**A Panopticon on My Wrist: The Biopower of Big Data Visualization for Wearables**
**K.J. Hepworth**

AUTHOR BIO K.J. Hepworth, Ph.D. is a communication design practitioner-researcher whose research explores the intersection between communication, power, and visualization. She is Associate Professor of Visual Journalism, Research Director of the Visualizing Science Project, and Co-Director of the Nevada Center for Data and Design at the Reynolds School of Journalism, University of Nevada, Reno, United States.
Khepworth@unr.edu

## Abstract

Big data visualization – the visual-spatial display of quantitative information culled from huge data sets – is now firmly embedded within the everyday experiences of people across the globe, yet scholarship on it remains surprisingly small. Within this literature, critical theorizations of big data visualizations are rare, as digital positivist perspectives dominate. This paper offers a critical, design-informed perspective on big data visualization in wearable health tracking ecosystems like FitBit. I argue that such visualizations are tools of individualized, neoliberal governance that operate largely through experiences of seduction and addiction to facilitate participation in the corporate capture and monetization of personal information. Exploration of my personal experience of the FitBit ecosystem illuminates this argument, and emphasizes the capacity for harm to individuals using these ecosystems, leading to an exploration of the complex professional challenges for user experience designers working on visualizations within the ecosystems of wearables.

## Introduction

In the early twenty-first century, a world without data visualization is basically unimaginable. Data visualizations are sense-making and decision-making tools commonly used countless times per day in myriad ways. Geographic information systems and environmental maps allow people to know where they are spatially, while schedules, timetables, and timelines do the same thing temporally. Infographics, interface and dashboard visualizations, and interactive charts make complex processes, subjects, and systems intelligible. People use myriad visualization formats to understand the world relationally (with network maps, tree diagrams, scatter plots, and Venn diagrams to name a few), and in comparison (for example, in bar charts, pie charts and line charts). Crucially, individuals, corporations, and governments depend on visualizations to navigate and interpret the seemingly ever-increasing quantities of personal behavioral and biometric data that activities now regularly generate; as epitomized by the quantified self movement and the pervasiveness of connected wearable devices such as fitness trackers and smart watches. This paper argues that visualizations associated with these devices constitute a dispersed, networked regulation of bodies; they encourage individual self-regulation as part of an extended neoliberal system of governance dependent on a kind of power called *biopower*.

**The Messiness of Big Data**

The data visualizations so many people experience daily, including those on wearables, are frequently driven by big data. This amorphous and looming concept appears somewhat abstract when publicly debated, despite it significantly impacting lived human experience due to its influence in the realms of consumption, social interaction, and public policy. The term 'big data' refers to three discrete things: specific ways of thinking about data and computation, data with specific qualities, and related analytical processes. In the first sense, the term relates to "a computational turn in thought and research" that favors the use of high volumes of available data (such as metadata generated from users' behavior on websites) and quantitative computational methods over all other approaches (Boyd & Crawford 2012, 665). The qualities that lead data to be considered 'big' include the speed of collection, the availability of real-time computational processing, low data accuracy, the correlation of data events across various data sources, and the volume of data compared with what was previously available in any given domain (Ekbia et al 2015, 1525-1526; Leszczynski 2015, 967). In terms of related analytical processes, big data is associated with methods for identifying patterns within and across large datasets, including automated data cleaning, machine learning, neural networks, and the predictive generation of new data (Salvo 2012, 37).

Low data accuracy is one quality of key concern for big data visualization. Big data routinely aggregates and conflates content with metadata, algorithmically calculated 'activity based intelligence' with actual human behavior, and digital profiles of users garnered from aggregated data sets with flesh and blood individuals. There is a potential for real-world damage to people from decisions made based on big data visualizations that rely on aggregated digital profiles (Crampon, 2015, p. 521; Poster, 1990, p. 126).

**Big Data Visualization Enables Human Decision Making**

Visualization is the necessary intermediary between big data and the human interpretation of that data. It mediates and filters data in such a way as to (hopefully) foster understanding and ideally, to result in effective, ethically-considered human decision making. The emphasis here is on *human* decision making because visualization is a uniquely human requirement. In the many rapidly automating and algorithm-dependent industries and sectors, computers and robots don't need a visual summary of data to base their decisions upon (Valle 2013, 2040). Despite our increasing collective dependence on big data, there are many fields of endeavor where decision making based on big data (through appraisal of visualizations) is deemed too important to outsource to computers, and consequently, we see the rapidly expanding offers of big data-driven visualizations available for the public.

While data visualization is essential to most knowledge domains in the twenty-first century, those fields of activity and decision making that are currently being transformed by big data – research, commerce, and government – have the most urgent need for, and dependence upon, visualization (Ali et al 2016, 656; Cook 2016, 135; Crampton 2015, 520). For researchers, visualization is both a means of analyzing research findings that is particularly crucial when dealing with big data, and an important way to share their research findings with a wider audience (Hepworth & Canon 2018, 53; Shoresh & Wong 2012, 5). For business purposes, visualization of aggregated behavioral and content production data is a means of identifying behaviors that can be monetized, price points that will be acceptable in given contexts, and managing supply chains (Salvo 2012, 39). For governments, visualization is critical to public policy decision making based on collecting big data at city, state, and national levels and across many areas, including environmental protection, national security, policing, and urban planning (Crampton 2015, 523; Lammerhirt 2015, 48; Leszczynski 2015, 968).

For individuals, as occupational and personal users and consumers of big data visualizations, interacting with them is a mandatory convenience. They offer benefits in terms of navigating our public and private worlds, both literally and figuratively, but these benefits are not a choice. While it is possible to greatly increase one's own exposure to big data visualizations, it is not possible to remove oneself from their influence entirely. Even the most self-sufficient, offline, and off-grid individual is monitored and regulated by countless visualizations used in government, commerce, and research.

The dramatic increase in the influence of big data analytics in many fields represents an attempt to better understand the world and the people in it "objectively," through the analysis of data that is perceived as more reliable than the human sources and subjects of that data (Thatcher et al 2016, 992). The meta-rationale behind big data visualization in all three spheres is that what can be collected, measured, and cross-referenced with other data sets can be managed more effectively. This 'numerical mediation', or the regulation and governance of individuals by the data trails they leave, is a twenty-first century technique of governmentality that is quintessentially American, and has been exported worldwide (Monea 2016).

Big data visualization can therefore be seen as a society-changing governance technology on par with the explosion of data visualization innovation that occurred in eighteenth- and nineteenth-century Europe. This geographical and temporal context saw both the advent of *biopower*, a term explained in greater detail below, and the invention of the most common chart types used today – the line chart, bar chart, and pie chart – as well as creation of standard visualization conventions within the fields of cartography, economics, geology, and statistics (Friendly 2008, 9; Schüll 2016, 326; Wainer 2005, 10). The timing and location of these innovations are inextricably linked to governance technologies; developed "in European countries at the height of their colonial expansion and industrial transformation... created to track demographics, trade, war, and debt; all the trappings of their growing empires" (Hepworth and Church 2018). Data visualization conventions are thus indelibly marked with the influence of colonialism, industrialization, and the expansion and increasing complexity of global capitalism.

**Big Data Visualization as a Neoliberal Technology**

Similarly, big data visualization is a product of its sociopolitical context: the global influence of neoliberalism, arguably the post-industrial successor to colonialism (von Sommaruga Howard 2016, 62). While neoliberalism has been conceived of as multiple overlapping phenomena – a dominant global ideology, a trend in processes and programs concerning capital and resources, and a shift in societal governance – this paper is concerned with the last (Springer 2012, 136-137). Sociologist and Foucauldian scholar Nikolas Rose identifies three key characteristics of neoliberal societies from this perspective, which includes an erosion of trust in expertise and political authority; the replacement of conceptions of the common good with individual self-government and social regulation; and lastly, the role of governments shifting to facilitate optimal self-government among their citizen-customers (Rose 1996, 57). Private and public spaces merge and blur, leading to an erosion of authentically public and private experiences. Under neoliberal influence, formerly public spaces, both physical and digital ones, become the property of corporate interests (either through government privatization or stealthy co-option), while private intellectual, physical, and physiological domains are also commercialized and regulated (Larner 2000, 8). Foucault referred to the regulation of bodily activities and processes as "biopower," and identified it as the primary regulatory concern of neoliberal governance (Foucault, 1979).

Central to neoliberal manifestations of biopower is the dispersed regulation of bodies, which encourage individual self-regulation of the internal processes of body and mind. In neoliberal manifestations of biopower, notions of agency, personal responsibility, and self-esteem are emphasized to such an extent that systemic, structural, and personal conditions are all but ignored (Cruikshank 1996, 245). Biopower has increasingly been used as a framework for understanding the regulatory effect of big data on lived human experience. In this vein, cultural anthropologist Natasha Dow Schüll refers to big data applications of this phenomenon as the "datafication of biopower" (2016, 328).

This paper views big data visualization through the dual lenses of Foucauldian neoliberalism and biopower, in order to critically situate it within its sociopolitical and historical contexts, particularly where such technologies have been dominant. It is worth noting that the use of big data visualizations in phones and wearables is not bounded by simple lines of economic disparity or privilege. Use of big data visualizations in phones and wearables is widespread within countries and communities not typically considered economically or technologically privileged; see for example research on Syrian refugees using smart phones for wayfinding in Europe (Gillespie et al 2018). Even people who don't regularly interact with wearables are nevertheless represented in and regulated by such visualizations, albeit at a distance. Neoliberal government and corporate policies aggregate data from online behavior and physical movement across locations with other biometric data, presenting them in visualizations for dashboards that are ultimately used for policy making and corporate decision making. A Foucauldian understanding of neoliberalism is used here because of its capacity to account for broad societal trends, such as the big current data revolution, as well as to effectively account for, and intersect with, mechanisms of power at a micro-level, such as the intimate workings of interaction experiences between data visualizations and individuals. It uses the visualizations designed for personal tracking devices as a case study, in concert with the metaphor of the Panopticon, to investigate how big data visualizations generally, and in the 'quantified self' movement specifically, function as technologies of neoliberal governance.

**The Quantified Self**

Along with the 'big data revolution' of the last ten years, public awareness of, and interest in, personal data aggregation (e.g., location mapping, biometric monitoring, activity tracking) has grown. The 'quantified self' movement is the latest chapter in the human fascination with self-knowledge and is the phenomena of individuals collecting data on themselves by intentionally tracking their own lives (Lupton 2013). While the quantified self movement encompasses individuals tracking a vast array of data about themselves for a wide range of purposes, several main motivations for personal tracking stand out: mood tracking, health and symptom tracking, and fitness activity tracking. Although experimental precursors to the quantified self movement date to the 1970s, it has become mainstream with the advent of affordable, appealing, big data collecting devices over the last decade, including the ubiquitous smart phone.

Today, mobile sensory technologies serve this human passion for self-documenting (Crampton 2015, 527). While the extent to which quantified self enthusiasts track their own data varies from individual to individual and from culture to culture, it is increasingly possible to track biological, behavioral, consumption, location, and social data in great detail, with the activity tracking code and sensors that come standard in commonplace technologies such as social media, websites, and smartphones. With this rise in personal data generation has come a recent proliferation of quantified self-oriented services – such as Daytum, Google Fit, and Strava – that allow individuals to add their own data from multiple sources and then view automated or semi-automated visualizations that allow them to look for patterns and trends within, and across, their own data sets.

Another group of popular consumer technologies, wrist-based personal trackers or wearables – for example Fitbit, Misfit, and Samsung Gear Fit, as well as a range of smart watches – are now commonplace personal tracking tools. While the early popularity of these devices was driven by their ability to collect physiological and activity data, recent sales indicate that today, the most popular trackers include integrated user experiences across multiple platforms, creating personal data 'ecosystems' that perform multiple functions beyond personal data collection. The term ecosystem is used across multiple disciplines, including user experience design, to capture the complexity and interdependence of coordinating and designing multiple individual platforms, media, interfaces, and products so that the experience of interacting with all of them is seamless, desirable, and entertaining for the end user. Personal data ecosystems surrounding wearables include an increasingly nuanced range of visualizations on the displays integrated into wearables; associated websites containing account information and more extensive visualizations; phone and tablet apps that offer yet more extensive visualizations of personal data (frequently comparing one's data with that of other users); email newsletters comparing one's activity to that of friends; and social media integrations, through which personal data is publicized, shared, and compared against other users of the same brand of wearable.

Visualized data about how our bodies function and move has become the latest addition to our curated online identities. As communications theorists Bossewich and Sinnreich argue, social capital "is increasingly constituted in the act of revelation, and in the methods by which we collect and reveal information to and about ourselves and others" (Bossewich and Sinnreich 2013, 225). As wrist-based personal tracking devices become increasingly widespread, so too does the urgency of critically framing their use as part of a broader neoliberal governance phenomena tied to the collection and analysis of big data. In the following sections, the governmental qualities of visualizations within wearable ecosystems are explored from a personal perspective, demonstrating the potential for harm intrinsic to the intersection between technologies of neoliberal governance and lived human experience.

**Seduction by Fitbit Visualizations**

My personal experience of interacting with the visualizations within the Fitbit ecosystem illustrates the seductiveness of visualizations within services catering to the quantified self. It should be noted that I am a queer, white, Australian, cis woman residing in the United States, and this demonstration of the designed operations of neoliberal capitalism upon my body is necessarily bounded by my own perspective.

I am entranced by my Fitbit data; like Narcissus captivated by his own reflection in a pool, I revel in the reflection of *me* represented in the Fitbit app's various visualizations. On my Fitbit Charge 2 and in the Fitbit phone app, a beating heart graphic delightfully animates in time to the beating of my actual heart. On the app, my exercise is tracked in a table that connects to a line chart of my heart rate during that activity. When I wake up, a sleep pattern bar chart tells me how long I slept, how well I slept, and it compares this to my sleep in previous weeks. The depictions of data about myself are embarrassingly thrilling. I find myself both eagerly awaiting the comparisons to my 'friends' with whose accounts I am connected within the Fitbit app and feeling a sense of shame at the level of pleasure this produces (Figure 1).

The easily comprehensible and apparently objective visualizations of my own physiological data are captivating, in part, because they utilize two rhetorical devices: enthymeme and pathos. An enthymeme is a syllogism whereby a key premise is left implicit in order to invite an audience "to participate in its own persuasion by filling in that unexpressed premise" (Blair 2004, 41). Data visualizations shape a persuasive visual argument through means of annotation, emphasis, summarizing, and visual hierarchy, but inevitably leave their users to draw conclusions from the data presented – conclusions which close the argument. This enthymeme quality of data visualizations is compounded by the ease with which data visualizations are understood, compared with text describing the same data (Johnson et al 2006, 5).

Their persuasive, apparent objectivity is furthered by the modernist-influenced functionalist aesthetic commonly found in visualizations, that uses "geometric layout, orderly typography, effective use of white space, and simple color composition" to give an aesthetic appearance of objectivity and efficiency (Shen 2018). The Fitbit ecosystem uses a gamified and cute version of these functionalist aesthetics. It adheres to the design qualities described above, but differs from standard functionalist visualizations by the use of friendly, rounded, humanist sans serif typefaces, rounded edges of data points, pill-shaped buttons, and the use of multiple, highly saturated colors. The effect of this cute functionalist aesthetic is apparent objectivity without intimidation: my personal data is rendered friendly and approachable at the same time it fosters certainty.

Extensively researched, user tested, and well-designed visualizations, such as those in the Fitbit personal tracking ecosystem, are even easier to understand because user experience designers intentionally accommodate users' cognitive, physical, and social factors, facilitating effective use of pathos, or emotional appeals to initially engage and maintain viewers' interest, and thereby increasing certainty of belief (Kostelnick 2007, 284; Skiba 2014, 268). Cognitive factors that well-designed data visualizations accommodate include mental models, motion tracking, pattern recognition, narrative appeals, and understanding of metaphor (Ali et al 2016, 656; Niepold et al 2008, 539; Zhang & Linn 2011, 1194). Physical factors include consideration of users' visual acuity, viewing distance, environmental context, and in interactive contexts, range of mobility (Reiner 2009, 360-361; Hepworth & Canon 2018, 59). Social factors that well-designed data visualizations accommodate include cultural associations of space and visual elements, and expertise in and previous exposure to represented data (Hepworth 2018, 516; Kostelnick 2007, 280, 284).

**Potential for Harm**

These visualizations of my own physiological data put me in a position of surveillance over my own body, giving me an impression that it is ordered, comprehensible, and amenable to continued optimization (Barton and Barton 1993, 146). Like Narcissus' reflection in the pool, the image of myself reflected in my visualized data has the power to harm. However, Narcissus and his reflected image were a more or less closed system in which his choices determined his fate, whereas me and my data are part of a much larger data ecosystem, constantly interacting with other users and corporations in the system. Our collected, aggregated, and visualized personal tracking data constitutes what media theorist Mark Poster refers to as 'the multiplication of the individual, the constitution of an additional self, one that may be acted upon to the detriment of the "real" self without that "real" self ever being aware of what is happening' (Poster 1990, 126). Writing in 1990, long before personal tracking became mainstream, Poster eloquently captured the potential for harm engendered within our personal data, and his argument has been echoed by other contemporary cultural theorists working on identity and surveillance (Cheney-Lippold, 2017; Koopman, 2019).

While none of these scholars write specifically about visualizations, their concerns hold true for visualization ecosystems, the data they represent, and their potential for harm. The potentially harmful effects of the user experiences of personal tracking devices include dark patterns, misrepresentation in visualizations, and misuse of the power imbalance inherent in this system (Bossewich and Sinnreich 2013, 226; Kuru and Forlizzi 2015, 490). Dark patterns refer to design features that encourage users to interact with interfaces in unintended ways or against their own interest: for example, when a pop up dialog box on a website or app contains 'accept' and 'cancel' buttons in the opposite order, and in opposite colors, to the standard layout, so that users who intend to 'accept' will inadvertently cancel, and vice versa.

The dark pattern quality of personal tracking apps takes advantage of the predictability of our cognitive processing strategies, in order to keep users engaged in the Fitbit ecosystem as often, and for as long as possible, through 'gamified' user experiences (Whitson 2013, 167). For example, in the Fitbit visualization in Figure 2, the use of saturated secondary colors in a horizontal bar chart format, and totaled numbers listed on the right, are both elements common in game leaderboard design that encourage comparison and the temptation to attempt to 'beat' one's prior record.

Another strategy used to encourage continued engagement is the visualization of certainty, presenting visualized personal data without depictions of margins of error or caveats. Personal trackers have a notoriously highly variable capacity to track the things that they present in neat, authoritative visualizations (Shcherbina et al 2017, 10). For example, Figure 2 depicts a Fitbit generated visualization of my sleep, despite Fitbit's inability to track actual sleep states; to do this it would need sensors attached to my scalp (Gilmore 2016, 2529). Instead, Fitbit predicts when I am in which sleep state based on pulse rate and movement. This is an example of what is referred to in commercial settings as 'activity based intelligence' and in security research as 'human dynamics', which are predictions of human behavior in the present and future based on partial data (Crampton 2015, 520).

Such predictive data creation – creating new, estimated data on individuals and groups of people from vast amounts of similar data – is a commonplace big data practice with potential to harm the real selves using this data to inform their decision making. As the sleep visualization example demonstrates, the predicted data may be highly inaccurate and, more importantly, it is rarely possible to verify where or how the visualizations that come from its analyses are generated. This opaque quality of big data has led several scholars to refer to it as 'dark data' (Ekbia et al 2015, 1539; Crampton 2015, 520). In the sleep visualization, I am presented with a compelling, authoritative chart of sleep patterns without any indication that the Fitbit is not actually capable of collecting such data. Such visualized estimations can have significant negative health impacts for people who are not aware of the device's margin of error and the predictive data processes used to collect the data informing these visualizations.

Fitbit also provides a chart of how my sleep compares with that of other women my age (Figure 3). This chart also appears without clarification or caveats and suggests a reassuring certainty. Looking at the chart, my first reaction was one of aspiration not just to comply but to excel. I had an unabashed desire to be *better* at sleep than other women my age. This visualization inspires comparison and competition in an activity that literally depends on unconsciousness, and raises several questions: What age range is being used to determine the 'my age' comparison? Which women are represented? They are the women who own Fitbits, and wear them to bed. Yet beyond gender and age, many factors affect quality of sleep, including medications, medical conditions, shift work, and stress. This chart breaks a golden rule of data visualization: to compare like with like. As such, the user experience encourages a mentality of compliance to an average that may not actually exist. It may not be possible for someone with my physiological make up and life situation to have what Fitbit has determined – using predicted data – is optimal sleep. In so doing, it presents a perfect neoliberal governance tool of self-regulation; if the goal is presented as attainable, but is not actually so, the activities undertaken to strive for it can be continuous, exhausting, and expensive. Absorbed in striving to optimize my activities as compared with fictional norms, I become a predictable, docile, and consumptive subject.

This line chart comparison of my personal data with that of other Fitbit users demonstrates a key quality of big data visualizations: they are almost completely automated, with other users' data being filtered through algorithms, middleware, and parallelized data processing before they are presented in the pre-set visualization template (Ali et al 2016, 656). Because, by definition, big data is data with low accuracy, a crucial job of automated and semi-automated visualization processes and tools is 'smoothing out' data, eliminating what is algorithmically identified as insignificant, and emphasizing what are algorithmically identified as potentially significant patterns.

This 'smoothing out' is an essential and helpful process in many areas of big data research, not just in visualization. However, it becomes troublesome when the process of 'smoothing out' – what is rendered significant and insignificant – is not made explicit to the users of visualization tools, processes, or outputs. For example, Fitbit's depicts false certainty across multiple visualizations that users cannot modify, combined with its use of proprietary algorithms, whose decision-making is both unknown and unalterable for users. Such opacity of visualization processes is referred to in the big data literatures as the 'black box problem,' and relates directly to 'dark data' (Ekbia et al 2015, 1539).

Such visualizations abandon the possible in favor of the unattainable. As sociologist Deborah Lupton observes, "when notions of health, wellbeing and productivity are produced via data drawn from self-monitoring, the social determinants of these attributes are hidden. Illness, emotional distress, lack of happiness, or lack of "productivity"… become represented primarily as failures of the individual" (Lupton 2013, 27-28). This presentation of data suggests, in a true neoliberal spirit, that results outside the mean are primarily due to individuals' poor self-control, goal setting, or inefficiency, and it encourages a compliance and aspiration mentality among users, sometimes referred to as "surveillant anxiety" (Horning 2014, Rose 1996, 59). Such visualizations thereby dissuade a consideration of the socio-political power structures, social contexts, and lived human experience behind the data (Fuchs 2017, 40).

**Power Imbalance**

While I wear my personal tracking device, Fitbit collects vast streams of activity data about my body. I have temporary access to a tiny fraction of this data, in templated visualizations over which I have no control. Fitbit determines which data I see, how long I have access to it, and who I am compared to in these visualizations. Fitbit, on the other hand, has access to all my data in perpetuity, and the benefit of completely customized cross-referencing of my data with other users' data to find trends and correlations over large numbers of users. Such power imbalances are often commented upon in big data literature, and in communication theory it has been studied for decades under the term 'knowledge gap' (Tichenor, Donohue & Olien 1980, 144; Ruckenstein & Schüll 2017).

**Neoliberal Panopticons**

French philosopher Michel Foucault famously used the metaphor of the Panopticon – Jeremy Bentham's much cited vision for the ideal prison (Foucault 1977, 14) – to describe the disciplinary influence of surveillance (Foucault 1980). In a culture where each person feels watched and judged — by their family, friends, and the organizations they are a part of — physical violence becomes a less crucial tool of governance than the regulatory influence of individual expectations, or self-surveillance. While all cultures have some level of social observation, Foucault associated the rise of extensive and subtle societal self-surveillance with the global rise of neoliberal governance from the late 1970s and onwards, leading to increasing reliance on biopower in both state and corporate governance.

**Networked, Personal Panopticons**

Foucault used the metaphor of the Panopticon to, among other things, demonstrate the power imbalance between government institutions and the people they were responsible for. However, given the neoliberal convergence of state and corporate power, the analogy is also appropriate to describing the power imbalances inherent to big data visualization generally and to the wearable context specifically. In this age of big data and social media, there is both a repetition and a fracturing of the disciplinary power structure, resulting in many, dispersed panopticons, like something akin to omnoptic surveillance (Jensen 2007; Mathiesen 1997). Each personal tracking device, and its associated companies and user experience ecosystems, can be thought of as mini, networked panopticons, "in which the many watch the many rather than just the few, as well as watching over themselves" (Gane 2012, 622). These complex networks of personal panopticons have been strikingly revealed in heatmap visualizations made publicly available by various personal tracking companies (see: https://www.strava.com/heatmap). In these cartographic visualizations, the location data of individual wearables is aggregated and represented as lines of travel with varying levels of brightness, providing a visual snapshot of the collective movement of everyone in the panopticon network, constituted by one wearable ecosystem (Langley, 2018).

One of the key differences between the nineteenth-century Panopticon and the present-day, networked, personal panopticons is that of motivation. As sociologist Nicholas Gane writes, their "techniques are seductive rather than coercive: no one is made to watch," although as has been discussed earlier, watching one's own data and that of others is compelled through a sophisticated harnessing of attention through gamification strategies and thoroughly researched and user tested visualizations (Gane 2012, 622). Our participation in this regulation is desire-driven, voluntary, and frequently, enthusiastic, as it provides a socially and personally valuable sense of self-care.

In my mini panopticon, the equivalent of the guard in the tower is a large, disparate surveillance team, any member of which may observe activity within the cells of my biometric and behavioral data at any time. This team includes myself, my friends, Fitbit staff, and any organization or individual who Fitbit sells my data to – these are sales I consented to when I signed the terms and conditions of using their services. The surveillance team expands to include advertisers, credit rating agencies, health insurance companies, and hospitals, among others, as they buy datasets containing my data (Ruckenstein & Schüll 2017).

Any member of this surveillance team can, and does, use this data to exert influence on my body and my lived experience. Sometimes, the surveillance team with regulatory influence extends beyond this already labyrinthine sprawl to include any viewer of certain visualizations of the panopticon network. Wearable ecosystem visualizations so regularly constitute harm by providing public access to personal data that there is a term for it: 'fit leaking.' For example, the Strava Heatmap was recently proven to constitute two kinds of inadvertent harm from fit leaking: there was a U.S. national security breach (caused by visualizing data of U.S. active duty military on secret U.S. bases) and the identification of individual users by name (identified through scraping location specific sections of data) (Hern, 2018, Langley, 2018). In terms of behavioral coercion, I attempt to increase or decrease activities based on the visualizations I see in the Fitbit app. My friends strike up conversations and sometimes competitive jokes based on our data shared in rankings, and these have a social regulatory effect on my behavior and my thought about my body. Fitbit uses my data to send me "encouraging" messages through my wearable to move in certain ways at certain times and to directly market premium, personal training subscription services. My one Fitbit thus constitutes a system of observation and judgement that ties in to the broader neoliberal governance trends in our society and has regulatory effects on how I think about, and relate to, my body.

In turn, I'm also the observer of my friends' biometric data through Fitbit's own social media network and through the data they share on other social media networks. Our personal tracking devices are connected with each other via social media features including "friends," they are connected with the companies that store their data, and to the organizations who buy our data from these companies. Personal tracking data is sold and re-sold, cross-correlated with other purchased data, (such as our spending habits, biometric data, location data, and social media activity) expanding the reach of the surveillance to which we willingly subject ourselves, eroding the private space in our lives, and having deleterious effects on our real selves (Elmer 2003). For example, one growth area of Fitbit's sales is in the 'employee wellness programs' market, where Fitbits are sold to companies and then distributed to employees (Till 2014, 452). The biometric data generated by employees in these programs goes to Fitbit, the employer, and their health insurance companies, as well as anyone else who purchases said data from Fitbit. Recent investigative reporting has uncovered substantial evidence of the data from personal trackers offered through such programs being used by health insurance companies not only to monitor participants, but also to penalize wearers whose data does not correspond with predetermined behavioral ideals. In some cases this has taken the form of dramatically increasing premiums, and in some extreme examples, denying health insurance (O'Neill 2018).

**User Experience Designers, Fun, and Agency**

User experience designers are crucial participants in the production of wearable device ecosystems. It is user experience designers who make interfaces and interactions not just usable, but also fun. They are essentially designing pleasurable interactions for elaborate neoliberal surveillance systems and creating visualizations that mislead and work against the interests of the wearers and users in multiple ways. Designing fun experiences that result in harm is a difficult professional dilemma. What user experience designers' ethical responsibilities in such contexts? And how much agency do user experience designers have to mitigate harm within their visualizations for wearables?

There is a pervasive thread in much design literature that challenges designers of all kinds to revolutionize a wide range of industries and societal processes. From Ken Garland's 1964 "First Things First Manifesto" to present day calls for design for the social good, designers have questioned their complicity in perpetuating consumerism. If I were to follow this vein, I might claim that user experience designers have a professional responsibility to disrupt wearable ecosystems and to inform users of the harm they are subjecting themselves to in ways more overt than the terms and conditions we all sign to use such services. Yet while calls for professional responsibility are important and necessary for advancing ethical practices, the strident nature of such calls in the design literature sometimes dramatically overestimate the professional agency of designers, and underestimate how regulated designers actually are, and consequently, how small their window of agency is.

A smaller body of literature, and one that incorporates perspectives on power and governance in designers' agency, argues that designers are bound into environmental, institutional, and knowledge discourses. These limit their agency far more than is commonly considered when design literature makes calls to action (Askehave & Zethsen 2003; Cairns 2002; Henderson 1991; Hepworth 2018). This power-informed literature amply demonstrates that, as well as constructing social processes, designers are also constructed by them: designers' thinking and action is bound by their training, their organizational affiliations, their environments, and tools. This is not to say that user experience designers lack agency, or that their practices cannot change, but meaningful efforts to mitigate harm in the user experience of visualizations in wearable ecosystems cannot rely solely on the efforts of individual user experience designers, or even on the field as a whole. Instead, approaches to mitigating harm in these ecosystems must cut across, as well as extend beyond, design disciplines.

**Ethical Pathways Forward**

The ethical challenge of big data visualization for development teams working on wearable ecosystems and beyond, is to honor and account for personal experiences and needs while also providing information and user experiences that minimize their potential harmful effects, insofar as this is possible within the prescribed bounds of culture and commerce in which they are situated. As described above, this work cannot fall solely in the hands of user experience designers. Rather, it requires the organizational recognition of, and concern for, the capacity for harm constituted within visualizations, and for the consequences of dark patterns being inadvertently incorporated into user experiences.

The persuasion inherent within big data visualizations can be kept within moral bounds by applying the same ethical rigor to data treatment practices and to designing visualizations as is applied to other areas of research. In terms of methodologies, the "ethical visualization workflow" has been developed and presented as one means of mitigating harm in data visualizations (Hepworth and Church 2019). This workflow calls for interdisciplinary collaboration and critical interrogation throughout all stages of data collection, interpretation, and visualization. While this workflow is proposed specifically for dealing with data in the digital humanities, it offers suggestions for a critical data handling and visualization practice that may have value for commercial design contexts. A key factor in increasing the ethical rigor and mitigating the harm in visualizations is interdisciplinary collaboration within teams and organizations. No one discipline, let alone a design team or an individual, can foresee all potential causes of harm in big data visualization. Collectively, however, it is possible to identify one another's professional and personal blind spots to produce visualizations that mitigate harm.

**Conclusion**

Personal tracking devices are presented by the companies who make them and the employers and health care providers who encourage their use, as tools of empowerment for individuals, "offering consumers a way to simultaneously embrace and outsource the task of lifestyle management" (Dow Schüll, 2016). The Fitbit user experience is marketed as "help[ing] people become more active, exercise more, [and] sleep better," by giving people the ability to track various physiological data points (Fitbit 2016). This feeds into the old management adage that you can't manage what you don't measure. But who is really managing whom? Personal tracking devices are no doubt empowering, in the ways Fitbit claims, for some people, some of the time. My own personal experience has been one of unabashed addiction and entertainment. But in concert with this personal empowerment, personal tracking devices also enable the unceasing encroachment of neoliberal governance and monetization via the datafication of biopower in formerly private bodily functions and spaces. The visualizations within personal tracking ecosystems are a crucial tool in this process, encouraging the continued use of, and delight in, personal tracking ecosystems, while also obscuring the complex systems of profit and surveillance underlying the promotion and broader social function of wearable technologies and the quantified self movement.

Since starting this paper, my infatuation with my Fitbit's reflection of me through visualizations of my data has, perhaps unsurprisingly, waned. I wish I could report that my growing awareness of the systems of surveillance that I willingly subject myself to was sufficient knowledge for me to cast aside my Fitbit. The truth is far more profane, and perhaps, more poignant. Two practical circumstances contributed to the end of my Narcissean romance with my visualized biometric data: the wristband on my Fitbit broke and my new phone is not compatible with my Fitbit. The appeal was not enough to compel me to purchase either a Fitbit-compatible smartphone or a new wristband. My agency as a user was limited to opt-in – and to accept all the privacy and ethical concerns, which, for a time, I was gladly willing to do – or opt-out. For all the attention we direct toward personal optimization, data accrual, and online sharing, we are nevertheless relentlessly, physically human: we are inconsequent, fallible, and subject to whims. Although the big data revolution and the visualizations stemming from it are subjecting us all to unprecedented surveillance, management, and dependence, we still have access to our direct experiences of our bodies and senses, which no amount of regulation or visualization can permanently obscure.

**Figures**

Figure 1. Reproduction of graphic of comparison with friends' steps, from Fitbit email correspondence. Artwork by the author.

Figure 2. Reproduction of Fitbit sleep data visualization. Artwork by the author.

Figure 3. Reproduction of Fitbit sleep state comparison bar chart. Artwork by the author.